\documentclass{article}
\usepackage{spconf,amsmath,graphicx,cite}
\usepackage{times}
\usepackage{amsbsy,color}
\usepackage{latexsym}
\usepackage{amssymb}
\usepackage{subfigure}
\usepackage{amsfonts}
\usepackage{epsfig}

\title{Distributed Low-Rank Estimation Based on Joint Iterative Optimization in Wireless Sensor Networks}

\name{ Songcen Xu$^*$, Rodrigo C. de Lamare$^\dag$ and H. Vincent Poor$^\maltese$
\vspace{-0.5em}}
{\address{$^*$Communications Research Group, Department of Electronics, University of York, U.K.\\
$^\dag$CETUC, PUC-Rio, Brazil / Dept. of Electronics, University of York, UK\\
$^\maltese$Dept. of Electrical Engineering, Princeton University, Princeton, NJ
08544, USA \\
Emails: sx520@york.ac.uk, rcdl500@york.ac.uk, poor@princeton.edu
\vspace{-0.75em}}}

\begin{document}
\maketitle

\begin{abstract}
This paper proposes a novel distributed reduced--rank scheme and an
adaptive algorithm for distributed estimation in wireless sensor
networks. The proposed distributed scheme is based on a
transformation that performs dimensionality reduction at each agent
of the network followed by a reduced-dimension parameter vector. A
distributed reduced-rank joint iterative estimation algorithm is
developed, which has the ability to achieve significantly reduced
communication overhead and improved performance when compared with
existing techniques. Simulation results illustrate the advantages of
the proposed strategy in terms of convergence rate and mean square
error performance.
\end{abstract}


\section{Introduction}
Distributed strategies have become fundamental for parameter
estimation in wireless networks and applications such as sensor
networks \cite{Lopes1,Lopes2,Xu,Xu1} and smart grids
\cite{Xie,Songcen}. Distributed processing techniques deal with the
extraction of information from data collected at nodes that are
distributed over a geographic area \cite{Lopes1}. In this context, a
specific node or agent in the network collects data from its
neighbors and combines them with its local information to generate
an improved estimate. However, when the unknown parameter vector to
be estimated has a large dimension, the network requires a large
communication bandwidth between neighbor nodes to transmit their
local estimate. This problem limits the application of existing
algorithms in applications with large data sets as the convergence
speed is dependent on the length of the parameter vector. Hence,
distributed dimensionality reduction has become an important tool
for distributed inference problems.

In order to perform dimensionality reduction, many algorithms have
been proposed in the literature, in the context of distributed
quantized Kalman Filtering \cite{Msechu,Xiao1}, quantized consensus
algorithms \cite{Pereira}, distributed principal subspace estimation
\cite{Li1}, single bit strategy \cite{Sayin} and Krylov subspaces
optimization techniques\cite{Chouvardas}. However, existing
algorithms are either too costly or have unsatisfactory performance
when processing a large number of parameters. As a result,
trade-offs between the amount of cooperation, communication and
system performance naturally exist. In this context, reduced--rank
techniques are powerful tools to perform dimensionality reduction,
which have been applied to DS--CDMA
system\cite{intspl,intcl,inttvt,ietcg,ccmmwf,wlmwf,jidfecho,Lamare,barc},
multi--input--multi--output (MIMO) equalization application
\cite{Lamare1}, spread--spectrum systems \cite{Lamare2} and
space--time interference suppression \cite{Lamare3,jiostap,ccmavf}.
However, limited research has been carried out on distributed
reduced-rank estimation. Related approaches to reduced-rank
techniques include compressive sensing-based strategies
\cite{Yao,stapjidf,l1stap,vfap,l1cg,alt}, which exploit sparsity to
reduce the number of parameters for estimation, and
attribute-distributed learning \cite{Zheng}, which employs agents
and a fusion center to meet the communication constraints.

In this paper, we propose a scheme for distributed signal processing
along with a distributed reduced-rank algorithm for parameter
estimation. In particular, the proposed algorithm is based on an
alternating optimization strategy and is called distributed
reduced-rank joint iterative optimization normalized least mean
squares (DRJIO--NLMS) algorithm. In contrast to prior work on
reduced-rank techniques and distributed methods, the proposed
reduced-rank strategy is distributed and performs dimensionality
reduction without costly decompositions at each agent. The proposed
DRJIO-NLMS algorithm is flexible with regards to the amount of
information that is exchanged, has low cost and high performance.
The transmitted information between each neighbor involves a
dimensionality reduction matrix and a reduced-dimension parameter
vector. The DRJIO-NLMS algorithm can also outperform competing
techniques.

This paper is organized as follows. Section 2 describes the system
model. In Section 3, the distributed dimensionality reduction and
adaptive processing scheme are introduced. The proposed distributed
reduced-rank algorithm is illustrated in Section 4. Simulation
results are provided in Section 5. Finally, we conclude the paper in
Section 6.

{\bf Notation}: We use boldface uppercase letters to denote matrices
and boldface lowercase letters to denote vectors. We use
$(\cdot)^{-1}$ to denote the inverse operator, $(\cdot)^H$ for
conjugate transposition and $(\cdot)^*$ for complex conjugate.

\section{System Model}

\begin{figure}[!htb]
\begin{center}
\def\epsfsize#1#2{0.9\columnwidth}
\epsfbox{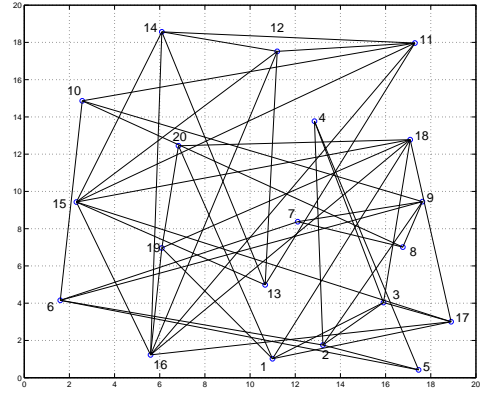} \vspace{-2.85em} \caption{\footnotesize
Network topology with 20 nodes}
\vskip -5pt
\label{fig1}
\end{center}
\end{figure}

A distributed wireless sensor network with N nodes, which have
limited processing capabilities, is considered with a partially
connected topology. A diffusion protocol is employed although other
strategies, such as incremental \cite{Lopes1} and consensus-based
\cite{Xie} could also be used. A partially connected network means
that nodes can exchange information only with their neighbors
determined by the connectivity topology. In contrast, a fully
connected network means that, data broadcast by a node can be
captured by all other nodes in the network \cite{Bertrand}. At every
time instant $i$, each node $k$ takes a scalar measurement $d_k(i)$
according to
\begin{equation}
{d_k(i)} = {\boldsymbol {\omega}}_0^H{\boldsymbol x_k(i)} +{n_k(i)},~~~ \label{desired signal}
i=1,2, \ldots, \textrm{N} ,
\end{equation}
where ${\boldsymbol x_k(i)}$ is the $M \times 1$ input signal
vector with zero mean and variance $\sigma_{x,k}^2$, ${ n_k(i)}$ is the noise sample at each node with zero mean
and variance $\sigma_{n,k}^2$. Through (\ref{desired signal}), we
can see that the measurements for all nodes are related to an
unknown parameter vector ${\boldsymbol {\omega}}_0$ with size $M
\times 1$, that would be estimated by the network. Fig.\ref{fig1} shows an example for a diffusion--type
wireless network with 20 nodes. The aim of such a network is to
compute an estimate of ${\boldsymbol{\omega}}_0$ in a distributed
fashion, which can minimize the cost function
\begin{equation}
{J_{\omega}({\boldsymbol \omega})} = {\mathbb{E}[ |{ d_k(i)}-
{\boldsymbol \omega}^H{\boldsymbol x_k(i)}|^2}] ,
\end{equation}
where $\mathbb{E}[\cdot]$ denotes the expectation operator. To solve
this problem, one possible technique is the adapt--then--combine
(ATC) diffusion strategy \cite{Lopes2}
\begin{equation}
\left\{\begin{array}{ll}
{\boldsymbol \psi}_k(i)= {\boldsymbol \omega}_k(i-1)+{\mu}_k {\boldsymbol x_k(i)}\big[{d_k(i)}-{\boldsymbol \omega}_k^H(i-1){\boldsymbol x_k(i)}\big]^*,\\
\ \\
{\boldsymbol {\omega}}_k(i)= \sum\limits_{l\in \mathcal{N}_k} c_{kl} \boldsymbol\psi_l(i),
\end{array}
\right.
\end{equation}
where $\mathcal{N}_k$ indicates the set of neighbors for node $k$,
$|\mathcal{N}_k|$ denotes the cardinality of $\mathcal{N}_k$ and
$c_{kl}$ is the combination coefficient, which is calculated under
the Metropolis rule
\begin{equation}
\left\{\begin{array}{ll}
c_{kl}= \frac{1}
{max(|\mathcal{N}_k|,|\mathcal{N}_l|)},\ \ \ \ \ \ \ \ \  \ \ \ \ \ \ \ \ \ \ \ \ \     $if\  $k\neq l$\  \ are\  linked$\\
c_{kl}=0,              \ \ \ \ \ \ \ \ \ \  \ \ \ \ \ \ \ \ \ \ \ \ \ \ \ \ \ \ \ \ \ \ \ \ \ \ \ \ \ \ \ \ \     $for\  $k$\  and\  $l$\ not\  linked$\\
c_{kk} = 1 - \sum\limits_{l\in \mathcal{N}_k / k} c_{kl}, \ \ \ \ \ \ \ \ \ \ \ \ \ \ \ \ \ \ \ \ \ $for\  $k$\ =\ $l$$
\end{array}
\right.
\end{equation}
and should satisfy
\begin{equation}
\sum\limits_{l} c_{kl} =1 , l\in \mathcal{N}_k \forall k .
\end{equation}
With this strategy, when the dimension of the unknown parameter
vector ${\boldsymbol {\omega}}_0$ is large, it could lead to a high
communication overhead between each neighbor node and the
convergence speed is reduced. In order to solve this problem and
optimize the distributed processing, we incorporate at the $k$th
node of the network distributed reduced--rank strategies based on
alternating optimization techniques.

\section{Distributed Dimensionality Reduction and Adaptive Processing}

\begin{figure}[!htb]
\begin{center}
\def\epsfsize#1#2{1.0\columnwidth}
\epsfbox{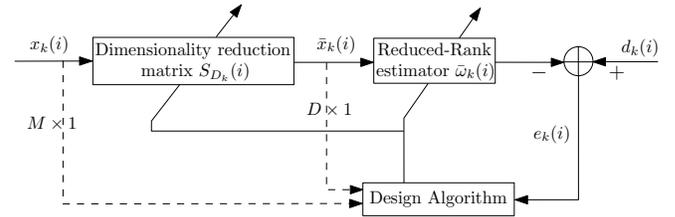} \vspace{-1.85em} \caption{\footnotesize
Proposed dimensionality reduction scheme at each node or agent.}
\label{fig2}
\vspace{-1.5em}
\end{center}
\end{figure}

The proposed distributed dimensionality reduction scheme, depicted
in Fig.\ref{fig2}, employs a transformation matrix $\boldsymbol
S_{D_k}(i)$ to process the input signal ${\boldsymbol x_k(i)}$ with
dimensions $M \times 1$ and projects it onto a lower $D \times 1$
dimensional subspace ${\bar{\boldsymbol x}_k(i)}$, where $D\ll M$.
Following this procedure, a reduced--rank estimator
$\bar{\boldsymbol\omega}_k(i)$ is computed, and the
$\bar{\boldsymbol\omega}_k(i)$ is transmitted through each neighbor
node. In particular, the transformation matrix $\boldsymbol
S_{D_k}(i)$ and reduced--rank estimator
$\bar{\boldsymbol\omega}_k(i)$ will be jointly optimized in the
proposed scheme according to the minimum mean--squared error (MMSE)
criterion.

Specifically, we start the description of the method with an
$M\times D$ matrix $\boldsymbol S_{D_k}(i)$, which carries out a
dimensionality reduction on the input signal as given by
\begin{equation}
\bar{\boldsymbol x}_k(i) = \boldsymbol S_{D_k}^H(i)\boldsymbol x_k(i),
\end{equation}
where, in what follows, all $D$--dimensional quantities are denoted
with a 'bar'. The design of $\boldsymbol S_{D_k}(i)$ and
$\bar{\boldsymbol\omega}_k(i)$ corresponds to the optimization
problem given by
\begin{equation}
\begin{split}
\hspace{-0.5em}\big\{ \boldsymbol S_{D_k}^{\rm opt},
\bar{\boldsymbol\omega}_k^{\rm opt} \big\} = \min_{\boldsymbol
S_{D_k}(i), \bar{\boldsymbol\omega}_k(i)}  & {\mathbb{E} [|{
d_k(i)}- {\boldsymbol{\bar{\omega}}_k}^H(i)\boldsymbol
S_{D_k}^H(i){\boldsymbol x_k(i)}|^2]}
\label{new cost}
\end{split}
\end{equation}
where $\boldsymbol{\bar{\omega}}_k(i)$ is the reduced--rank
estimator. By fixing $\boldsymbol S_{D_k}(i)$ and minimizing
(\ref{new cost}) with respect to $\boldsymbol{\bar{\omega}}_k(i)$,
we have
\begin{equation}
\boldsymbol{\bar{\omega}}_k(i)= \bar{\boldsymbol R}_k^{-1}(i)\bar{\boldsymbol p}_k(i),\label{wk}
\end{equation}
where $\bar{\boldsymbol R}_k(i)=\mathbb{E}[\boldsymbol
S_{D_k}^H(i)\boldsymbol x_k(i)\boldsymbol x_k^H(i)\boldsymbol
S_{D_k}(i)]=\mathbb{E}[\bar{\boldsymbol x}_k(i)\bar{\boldsymbol
x}_k^H(i)]$ and $\bar{\boldsymbol
p}_k(i)=\mathbb{E}[d_k^*(i)\boldsymbol S_{D_k}^H(i)\boldsymbol
x_k(i)]=\mathbb{E}[d_k^*(i)\bar{\boldsymbol x}_k(i)]$. We then fix
$\boldsymbol{\bar{\omega}}_k(i)$ and minimize (\ref{new cost}) with
respect to $\boldsymbol S_{D_k}(i)$, and arrive at the following
expression
\begin{equation}
\boldsymbol S_{D_k}(i)= \boldsymbol R_k^{-1}(i)\boldsymbol P_{D_k}(i)\bar{\boldsymbol R}_{\bar{\boldsymbol\omega}_k}^{-1}(i),\label{sdk}
\end{equation}
where $\boldsymbol R_k(i)=\mathbb{E}[\boldsymbol x_k(i)\boldsymbol x_k^H(i)]$, $\boldsymbol P_{D_k}(i)=\mathbb{E}[d_k^*(i)\boldsymbol x_k(i)\bar{\boldsymbol\omega}_k^H(i)]$ and $\bar{\boldsymbol R}_{\bar{\boldsymbol\omega}_k}(i)=\mathbb{E}[\boldsymbol{\bar{\omega}}_k(i)\boldsymbol{\bar{\omega}}_k^H(i)]$. At this stage, the associated reduced--rank MMSE is described as
\begin{equation}
\mathbf{MMSE} = \sigma_{d_k}^2-\bar{\boldsymbol p}_k^H(i)\bar{\boldsymbol R}_k^{-1}(i)\bar{\boldsymbol p}_k(i)
\end{equation}
where $\sigma_{d_k}^2=\mathbb{E}[|d_k(i)|^2]$. Because there is no
closed-form expression for $\boldsymbol S_{D_k}(i)$ and
$\bar{\boldsymbol\omega}_k(i)$ as they depend on each, we need a
strategy to compute the parameters. The proposed strategy is based
on an alternating optimization of  $\boldsymbol S_{D_k}(i)$ and
$\bar{\boldsymbol\omega}_k(i)$. In the next section, we develop a
distributed reduced-rank algorithm to compute the parameters of
interest.

\section{Proposed Distributed Reduced-Rank Algorithm}

In this section, we present the proposed distributed reduced--rank
algorithm for distributed estimation, namely DRJIO--NLMS. Unlike
prior work \cite{Li1,Sayin,Chouvardas,Rcdl3,smce}, the proposed
algorithm does \textbf{NOT} require
\begin{itemize}
\item an $M\times M$ auto--correlation matrix of the input signal and an $M\times 1$ cross--correlation vector between the input signal and the desired signal used to build the Krylov subspace \cite{Chouvardas}
\item Additional cost to perform eigen--decompositions \cite{Li1}
\item Extra adaptive processing at the local node \cite{Sayin}
\item Costly convex optimization at the local node, which introduces extra complexity \cite{Chouvardas}.
\end{itemize}
The DRJIO-NLMS algorithm is flexible, has a low cost and a very fast
convergence speed. The proposed DRJIO-NLMS algorithm needs to
optimize the parameters in (7) jointly. Therefore, it solves the
following optimization problem in an alternating fashion:
\begin{equation}
\begin{split}
\big[ \boldsymbol S_{D_{k}}^{\rm opt},
\boldsymbol{\bar{\omega}}_{k}^{\rm opt} \big] & = \min_{\boldsymbol
S_{D_k}(i), \bar{\boldsymbol\omega}_k^H(i)}
||{\boldsymbol{\bar{\omega}}_k}(i) -
{\boldsymbol{\bar{\omega}}_k}(i-1)||^2 \\ & \quad ~~~~~~ +||\boldsymbol S_{D_k}(i)-\boldsymbol S_{D_k}(i-1)||^2 \\
{\rm subject} ~{\rm to}~~ &
\bar{\boldsymbol\omega}_k^H(i)\boldsymbol S_{D_k}^H(i){\boldsymbol
x_k(i)} = d_k(i). \label{opt}
\end{split}
\end{equation}
Using the method of Lagrange multipliers and considering
$\{\boldsymbol S_{D_{k}}, \boldsymbol{\bar{\omega}}_{k}\}$ jointly,
we arrive at the
following Lagrangian:
\begin{equation}
\begin{split}
\mathcal{L}_k &= ||{\boldsymbol{\bar{\omega}}_k}(i)-{\boldsymbol{\bar{\omega}}_k}(i-1)||^2+||\boldsymbol S_{D_k}(i)-\boldsymbol S_{D_k}(i-1)||^2 \\
&\ \ \ \ +\mathfrak{R}[\lambda_1^*(d_k(i)-\bar{\boldsymbol\omega}_k^H(i)\boldsymbol S_{D_k}^H(i-1){\boldsymbol x_k(i)})]\\
&\ \ \ \ +\mathfrak{R}[\lambda_2^*(d_k(i)-\bar{\boldsymbol\omega}_k^H(i-1)\boldsymbol S_{D_k}^H(i){\boldsymbol x_k(i)})], \label{cost funcotion}
\end{split}
\end{equation}
where , $\lambda_1$, $\lambda_2$ are scalar Lagrange multipliers,
$||\cdot||$ denotes the Frobenius norm, and the operator
$\mathfrak{R}[\cdot]$ retains the real part of the argument. By
computing the gradient terms of (\ref{cost funcotion}) with respect
to $\boldsymbol{\bar{\omega}}_k(i)$, $\boldsymbol S_{D_k}(i)$,
$\lambda_1$ and $\lambda_2$, respectively, we obtain
\begin{equation}
\nabla_{\boldsymbol{\bar{\omega}}_k(i)}\mathcal{L}=2\big(\boldsymbol{\bar{\omega}}_k(i)-\boldsymbol{\bar{\omega}}_k(i-1)\big)+\boldsymbol S_{D_k}^H(i-1)\boldsymbol x_k(i)\lambda_1\label{a1}
\end{equation}
\begin{equation}
\nabla_{\boldsymbol S_{D_k}(i)}\mathcal{L}=2\big(\boldsymbol S_{D_k}(i)-\boldsymbol S_{D_k}(i-1)\big)+\boldsymbol x_k(i)\boldsymbol{\bar{\omega}}_k(i-1)\lambda_2
\end{equation}
\begin{equation}
\nabla_{\lambda_1}\mathcal{L}=d_k(i)-\boldsymbol{\bar{\omega}}^H_k(i)\boldsymbol S_{D_k}^H(i-1)\boldsymbol x_k(i)
\end{equation}
\begin{equation}
\nabla_{\lambda_2}\mathcal{L}=d_k(i)-\boldsymbol{\bar{\omega}}_k^H(i-1)\boldsymbol S_{D_k}^H(i)\boldsymbol x_k(i).\label{a2}
\end{equation}
By setting (\ref{a1})--(\ref{a2}) to zero and solving the remaining
equations, we obtain the recursions of the proposed DRJIO--NLMS
algorithm described by
\begin{equation}
\bar{\boldsymbol\omega}_k(i)=\bar{\boldsymbol\omega}_k(i-1)+\mu(i)e_k^*(i)\bar{\boldsymbol x}_k(i)
\end{equation}
\begin{equation}
\boldsymbol S_{D_k}(i)=\boldsymbol S_{D_k}(i-1)+\eta(i)e_k^*(i)\boldsymbol x_k(i)\bar{\boldsymbol\omega}_k^H(i-1)
\end{equation}
where $e_k(i)=d_k(i)-\bar{\boldsymbol\omega}_k^H(i-1)\boldsymbol
S_{D_k}^H(i-1){\boldsymbol x_k(i)}$, $\mu(i)=\frac{\mu_0}{\boldsymbol
x_k^H(i)\boldsymbol x_k(i)}$ and
$\eta(i)=\frac{\eta_0}{\bar{\boldsymbol\omega}_k^H(i-1)\bar{\boldsymbol\omega}_k(i-1)\boldsymbol
x_k^H(i)\boldsymbol x_k(i)}$ are the time--varying step sizes, while
$\mu_0$ and $\eta_0$ are the convergence factors. The recursions are
computed in an alternating way with one iteration per time instant
at each node.

The proposed DRJIO--NLMS algorithm includes two steps, namely
adaptation step and combination step which are performed in an
alternating procedure which is detailed next.
\begin{itemize}
\item Adaptation step
\end{itemize}
For the adaptation step, at each time instant $i$=1,2, . . . , I,
each node $k$=1,2, \ldots, N, starts from generating a local
reduced--rank estimator through
\begin{equation}
{\boldsymbol {\bar{\psi}}}_k(i)= \bar{\boldsymbol\omega}_k(i-1)+\mu(i)e_k^*(i)\bar{\boldsymbol x}_k(i),
\end{equation}
where $e_k(i)=d_k(i)-\bar{\boldsymbol\omega}_k^H(i-1)\boldsymbol S_{D_k}^H(i){\boldsymbol x_k(i)}$. This local reduced--rank estimator ${\boldsymbol {\bar{\psi}}}_k(i)$ will be transmitted to all its neighbor nodes under the network topology structure.

Then, each node $k$=1,2, \ldots, N, will update its dimensionality reduction matrix according to
\begin{equation}
\boldsymbol S_{D_k}(i)=\boldsymbol S_{D_k}(i-1)+\eta(i)e_k^*(i)\boldsymbol x_k(i)\bar{\boldsymbol\omega}_k(i-1),
\end{equation}
and keep it locally.
\begin{itemize}
\item Combination step
\end{itemize}
At each time instant $i$=1,2, . . . , I, the combination step starts
after the adaptation step finishes. Each node will combine the local
reduced--rank estimators from its neighbor nodes and itself through
\begin{equation}
\bar{{\boldsymbol {\omega}}}_k(i)= \sum\limits_{l\in \mathcal{N}_k} c_{kl}\boldsymbol{\bar{\psi}}_l(i),
\end{equation}
to compute the reduced--rank estimator $\bar{\boldsymbol
{\omega}}_k(i)$.

After the final iteration $I$, each node will generate the full--rank estimator ${\boldsymbol {\omega}}_k(I)$ from
\begin{equation}
{\boldsymbol {\omega}}_k(I)= \boldsymbol S_{D_k}(I){\boldsymbol {\bar{\omega}}}_k(I).
\end{equation}
In conclusion, during the distributed processing steps, only the
local reduced--rank estimator ${\boldsymbol {\bar{\psi}}}_k(i)$ will
be transmitted through the network. The proposed DRJIO--NLMS
algorithm is detailed in Table.\ref{table1}.

\begin{table}[!htb]\small
\vspace{-1.5em}
\centering \caption{{The DRJIO--NLMS Algorithm}}
\begin{tabular}{l}\hline
Initialize: ${\boldsymbol {\bar{\omega}}}_k(0)$=0\\
For each time instant $i$=1,2, . . . , I\\
\ \ \ \ For each node $k$=1,2, \ldots, N\\
\ \ \ \ \ \ \ \ \ \ ${\boldsymbol {\bar{\psi}}}_k(i)= \bar{\boldsymbol\omega}_k(i-1)+\mu(i)e_k^*(i)\bar{\boldsymbol x}_k(i)$\\
\ \ \ \ \ \ \ \ \ \ where $e_k(i)=d_k(i)-\bar{\boldsymbol\omega}_k^H(i-1)\boldsymbol S_{D_k}^H(i){\boldsymbol x_k(i)}$\\
\ \ \ \ \ \ \ \ \ \ \ \% ${\boldsymbol {\bar{\psi}}}_k(i)$ is the local reduced--rank estimator and will be \\
\ \ \ \ \ \ \ \ \ \ \ \% sent to all neighbor nodes of node $k$ under the network \\
\ \ \ \ \ \ \ \ \ \ \ \%  topology structure.\\
\ \ \ \ \ \ \ \ \ \ $\boldsymbol S_{D_k}(i)=\boldsymbol S_{D_k}(i-1)+\eta(i)e_k^*(i)\boldsymbol x_k(i)\bar{\boldsymbol\omega}_k(i-1)$\\
\ \ \ \ \ \ \ \ \ \ \% The dimensionality reduction matrix $\boldsymbol S_{D_k}(i)$  \\
\ \ \ \ \ \ \ \ \ \ \% will be updated and kept locally.\\
\ \ \ \ end\\
\ \ \ \ For each node $k$=1,2, \ldots, N\\
\ \ \ \ \ \ \ \ \ \ $\bar{{\boldsymbol {\omega}}}_k(i)= \sum\limits_{l\in \mathcal{N}_k} c_{kl}\boldsymbol{\bar{\psi}}_l(i)$\\
\ \ \ \ \ \ \ \ \ \ \% The reduced--rank estimator $\bar{{\boldsymbol {\omega}}}_k(i)$ \\
\ \ \ \ \ \ \ \ \ \ \% will be updated and kept locally.\\

\ \ \ \ end\\
end\\
After the final iteration $I$\\
For each node $k$=1,2, \ldots, N\\
\ \ \ \ ${\boldsymbol {\omega}}_k(I)= \boldsymbol S_{D_k}(I){\boldsymbol {\bar{\omega}}}_k(I)$ \\
\ \ \ \ where ${\boldsymbol {\omega}}_k(I)$ is the final full--rank estimator.\\
end\\
\hline
\end{tabular}
\label{table1}
\vspace{-1.5em}
\end{table}

The computational complexity of the proposed DRJIO--NLMS algorithm
is $O(DM)$. The distributed NLMS algorithm has a complexity $O(M)$,
while the complexity of the distributed recursive least squares
(RLS) algorithm \cite{Cattivelli} is $O(M^2)$. For both the Krylov
Subspace NLMS \cite{Chouvardas} and distributed principal subspace
estimation algorithms \cite{Li1}, the complexity reaches $O(M^3)$.
Thus, the proposed DRJIO--NLMS algorithm has a much lower
computational complexity, and because $D\ll M$, it is as simple as
the distributed NLMS algorithm. In addition, the dimensionality
reduction results in a decrease in the number of transmitted
parameters from $M$ to $D$ which corresponds to a less stringent
bandwidth requirement.

\section{Simulation results}
In this section, we compare our proposed DRJIO--NLMS algorithm with
the distributed NLMS algorithm, distributed RLS algorithm
\cite{Cattivelli}, Krylov subspace NLMS \cite{Chouvardas} and
distributed principal subspace estimation \cite{Li1}, based on their
mean-squared error (MSE) performance. With the network topology
structure outlined in Fig. \ref{fig1} with $N=20$ nodes, we consider
numerical simulations under three scenarios
\begin{itemize}
\item Full--rank system with $M$=20
\item Sparse system with $M$=20 ($D$ valid coefficients and $M-D$ zeros coefficients)
\item Full--rank system with $M$=60
\end{itemize}
The input signal is generated as ${\boldsymbol x_k(i)}=[x_k(i)\ \ \
x_k(i-1)\ \ \ ...\ \ \ x_k(i-M+1)]$  and
$x_k(i)=u_k(i)+\alpha_kx_k(i-1)$, where $\alpha_k$ is a correlation
coefficient and $u_k(i)$ is a white noise process with variance
$\sigma^2_{u,k}= 1-|\alpha_k|^2$, to ensure the variance of
${\boldsymbol x_k(i)}$ is $\sigma^2_{x,k}= 1$. The noise samples are
modeled as complex Gaussian noise with variance of $\sigma^2_{n,k}=
0.001$. We assume that the network has perfect transmission between
linked nodes.

\begin{figure}[!htb]
\begin{center}
\def\epsfsize#1#2{1.0\columnwidth}
\epsfbox{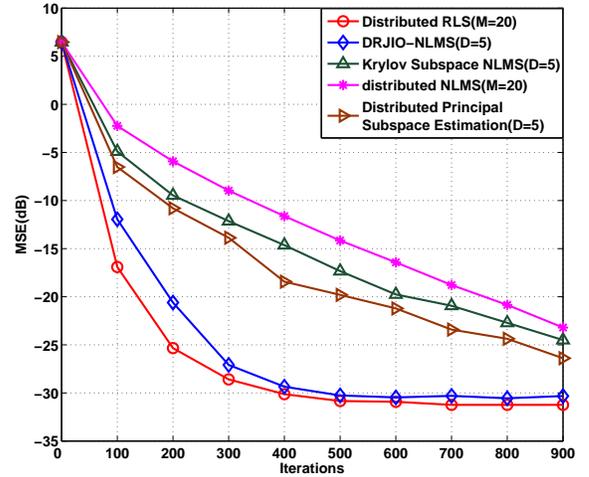} \vspace{-1.95em} \caption{\footnotesize
Full--rank system with $M$=20}  \label{fig3}
\vspace{-1.5em}
\end{center}
\end{figure}

The step size $\mu_0$ for the distributed NLMS algorithm, Krylov
subspace NLMS, distributed principal subspace estimation and
DRJIO--NLMS is 0.15 and the $\eta_0$ is set to 0.5. For the
distributed RLS algorithm, the forgetting factor $\lambda$ is equal
to 0.99 and the $\delta$ is 0.11. In Fig. \ref{fig3}, we compare the
proposed DRJIO--NLMS with the existing strategies using the
full--rank system with $M$=20 and $D$=5. The dimensionality
reduction matrix $\boldsymbol S_{D_k}(0)$ is initialized as
$[\boldsymbol I_D \ \ \boldsymbol 0_{D,M-D}]^T$. We observe that the
proposed DRJIO--NLMS algorithm has a better performance on both the
MSE level and convergence rate, which is very close to the
distributed RLS algorithm. However, its complexity is an order of
magnitude lower than the distributed RLS algorithm.

\begin{figure}[!htb]
\begin{center}
\def\epsfsize#1#2{1.0\columnwidth}
\epsfbox{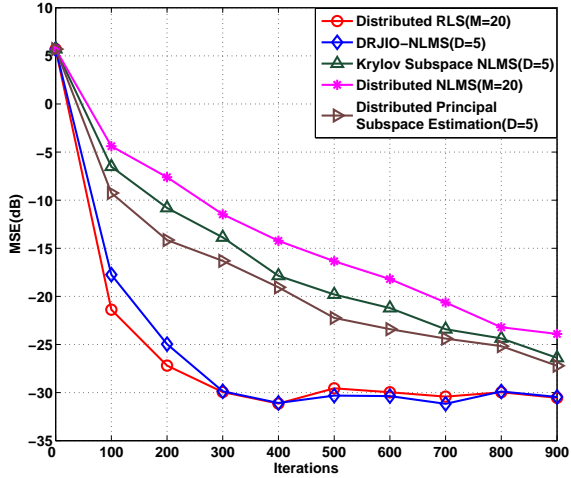} \vspace{-1.95em} \caption{\footnotesize Sparse
system with $M$=20} \vskip -5pt \label{fig4}
\end{center}
\end{figure}

\begin{figure}[!htb]
\begin{center}
\def\epsfsize#1#2{1.0\columnwidth}
\epsfbox{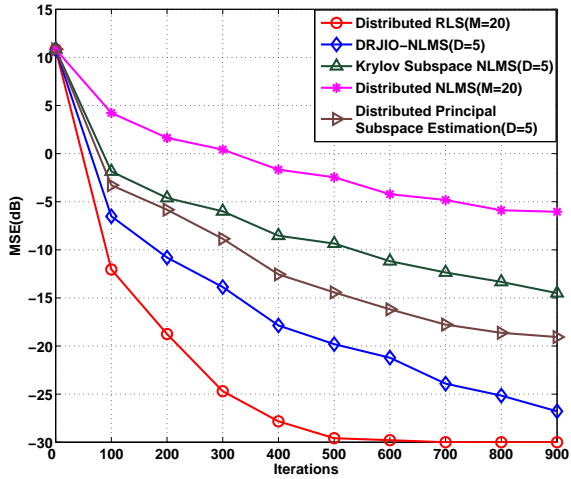} \vspace{-1.95em} \caption{\footnotesize
Full--rank system with $M$=60} \vskip -5pt \label{fig5}
\end{center}
\end{figure}

In a sparse system with $M$=20 scenario, the convergence rate for
all algorithms increase, and the proposed DRJIO--NLMS algorithm
still has an excellent performance as shown in Fig. \ref{fig4}.
Specifically, the proposed DRJIO--NLMS algorithm performs very close
to the distributed RLS algorithm and outperforms to other analyzed
algorithms. When the full--rank system $M$ increases to 60, Fig.
\ref{fig5} illustrates that, the proposed DRJIO--NLMS algorithm also
shows a very fast convergence rate. For the distributed NLMS
algorithm, Krylov subspace NLMS and distributed principal subspace
estimation, their convergence speed is much lower.

\section{Conclusion}

In this paper, we have proposed a novel distributed reduced--rank
scheme along with an efficient algorithm for distributed estimation
in wireless sensor networks. Simulation results have shown that the
proposed DRJIO-NLMS algorithm has better performance and lower cost
than existing algorithms in all the three scenarios considered.
Furthermore, the proposed scheme requires the transmission of only
$D$ parameters instead of $M$, resulting in higher bandwidth
efficiency than standard schemes.


{\footnotesize
\bibliographystyle{IEEEbib}
\bibliography{reference}}

\begin{thebibliography}{10}

\bibitem{Lopes1}
C.~G. Lopes and A.~H. Sayed,
\newblock ``Incremental adaptive strategies over distributed networks,''
\newblock {\em IEEE Trans. Signal Process.}, vol. 48, no. 8, pp. 223--229, Aug
  2007.

\bibitem{Lopes2}
C.~G. Lopes and A.~H. Sayed,
\newblock ``Diffusion least-mean squares over adaptive networks: Formulation
  and performance analysis,''
\newblock {\em IEEE Trans. Signal Process.}, vol. 56, no. 7, pp. 3122--3136,
  July 2008.

\bibitem{Xu}
S.~Xu and R.~C. de~Lamare,
\newblock ``Distributed conjugate gradient strategies for distributed
  estimation over sensor networks,''
\newblock {\em Proc. Sensor Signal Processing for Defence 2012}, London, UK,
  2012.

\bibitem{Xu1}
S.~Xu, R.~C. de~Lamare, and H.~V. Poor,
\newblock ``Adaptive link selection strategies for distributed estimation in
  diffusion wireless networks,''
\newblock {\em Proc. IEEE International Conference on Acoustics, Speech, and
  Signal Processing}, Vancouver, Canada 2013.

\bibitem{Xie}
L.~Xie, D.-H. Choi, S.~Kar, and H.~V. Poor,
\newblock ``Fully distributed state estimation for wide-area monitoring
  systems,''
\newblock {\em IEEE Trans. Smart Grid}, vol. 3, no. 3, pp. 1154--1169,
  September 2012.

\bibitem{Songcen}
Songcen Xu, R.C. de~Lamare, and H.V. Poor,
\newblock ``Dynamic topology adaptation for distributed estimation in smart
  grids,''
\newblock {\em Proc. IEEE International Workshop on Computational Advances in
  Multi-Sensor Adaptive Processing}, pp. 420--423, Saint Martin, Dec 2013.

\bibitem{Msechu}
E.~Msechu, S.~Roumeliotis, A.~Ribeiro, and G.~Giannakis,
\newblock ``Decentralized quantized kalman filtering with scalable
  communication cost,''
\newblock {\em IEEE Trans. Signal Process.}, vol. 56, no. 8, pp. 3727–3741,
  October 2008.

\bibitem{Xiao1}
J.-J. Xiao, A.~Ribeiro, Z.-Q. Luo, and G.~Giannakis,
\newblock ``Decentralized quantized kalman filtering with scalable
  communication cost,''
\newblock {\em IEEE Signal Process. Mag.}, vol. 23, no. 4, pp. 27--41, July
  2006.

\bibitem{Pereira}
S.~Pereira and A.~Pages-Zamora,
\newblock ``Distributed consensus in wireless sensor networks with quantized
  information exchange,''
\newblock {\em Proc. IEEE 9th Workshop Signal Process. Adv. Wireless
  Communication}, pp. 241--245, July 2008.

\bibitem{Li1}
Lin Li, A.~Scaglione, and J.H. Manton,
\newblock ``Distributed principal subspace estimation in wireless sensor
  networks,''
\newblock {\em IEEE Journal of Selected Topics in Signal Processing}, vol. 5,
  no. 4, pp. 725--738, 2011.

\bibitem{Sayin}
M.O. Sayin and S.S. Kozat,
\newblock ``Single bit and reduced dimension diffusion strategies over
  distributed networks,''
\newblock {\em IEEE Signal Process. Lett.}, vol. 20, no. 10, pp. 976--979,
  2013.

\bibitem{Chouvardas}
S.~Chouvardas, K.~Slavakis, and S.~Theodoridis,
\newblock ``Trading off complexity with communication costs in distributed
  adaptive learning via krylov subspaces for dimensionality reduction,''
\newblock {\em IEEE Journal of Selected Topics in Signal Processing}, vol. 7,
  no. 2, pp. 257--273, 2013.

\bibitem{intspl}
R.C. de~Lamare and R.~Sampaio-Neto,
\newblock ``Adaptive reduced-rank mmse filtering with interpolated fir filters
  and adaptive interpolators,''
\newblock {\em Signal Processing Letters, IEEE}, vol. 12, no. 3, pp. 177--180,
  March 2005.

\bibitem{intcl}
R.C. de~Lamare and R.~Sampaio-Neto,
\newblock ``Reduced-rank interference suppression for ds-cdma based on
  interpolated fir filters,''
\newblock {\em Communications Letters, IEEE}, vol. 9, no. 3, pp. 213--215,
  March 2005.

\bibitem{inttvt}
R.C. de~Lamare and R.~Sampaio-Neto,
\newblock ``Adaptive interference suppression for ds-cdma systems based on
  interpolated fir filters with adaptive interpolators in multipath channels,''
\newblock {\em Vehicular Technology, IEEE Transactions on}, vol. 56, no. 5, pp.
  2457--2474, Sept 2007.

\bibitem{ietcg}
L.~Wang and R.C. de~Lamare,
\newblock ``Constrained adaptive filtering algorithms based on conjugate
  gradient techniques for beamforming,''
\newblock {\em Signal Processing, IET}, vol. 4, no. 6, pp. 686--697, Dec 2010.

\bibitem{ccmmwf}
R.C. de~Lamare, M.~Haardt, and R.~Sampaio-Neto,
\newblock ``Blind adaptive constrained reduced-rank parameter estimation based
  on constant modulus design for cdma interference suppression,''
\newblock {\em Signal Processing, IEEE Transactions on}, vol. 56, no. 6, pp.
  2470--2482, June 2008.

\bibitem{wlmwf}
Nuan Song, R.C. de~Lamare, M.~Haardt, and M.~Wolf,
\newblock ``Adaptive widely linear reduced-rank interference suppression based
  on the multistage wiener filter,''
\newblock {\em Signal Processing, IEEE Transactions on}, vol. 60, no. 8, pp.
  4003--4016, Aug 2012.

\bibitem{jidfecho}
M.~Yukawa, R.C. de~Lamare, and R.~Sampaio-Neto,
\newblock ``Efficient acoustic echo cancellation with reduced-rank adaptive
  filtering based on selective decimation and adaptive interpolation,''
\newblock {\em Audio, Speech, and Language Processing, IEEE Transactions on},
  vol. 16, no. 4, pp. 696--710, May 2008.

\bibitem{Lamare}
R.C.~De Lamare and R.~Sampaio-Neto,
\newblock ``Reduced--rank adaptive filtering based on joint iterative
  optimization of adaptive filters,''
\newblock {\em IEEE Signal Process. Lett.}, vol. 14, no. 12, pp. 980--983,
  2007.

\bibitem{barc}
R.C. de~Lamare, R.~Sampaio-Neto, and M.~Haardt,
\newblock ``Blind adaptive constrained constant-modulus reduced-rank
  interference suppression algorithms based on interpolation and switched
  decimation,''
\newblock {\em Signal Processing, IEEE Transactions on}, vol. 59, no. 2, pp.
  681--695, Feb 2011.

\bibitem{Lamare1}
R.C.~De Lamare and R.~Sampaio-Neto,
\newblock ``Adaptive reduced-rank equalization algorithms based on alternating
  optimization design techniques for mimo systems,''
\newblock {\em IEEE Trans. Vehi. Techn.}, vol. 60, no. 6, pp. 2482--2494, 2011.

\bibitem{Lamare2}
R.C.~De Lamare and R.~Sampaio-Neto,
\newblock ``Reduced-rank space--time adaptive interference suppression with
  joint iterative least squares algorithms for spread-spectrum systems,''
\newblock {\em IEEE Trans. Vehi. Techn.}, vol. 59, no. 3, pp. 1217--1228, 2010.

\bibitem{Lamare3}
R.C.~De Lamare and R.~Sampaio-Neto,
\newblock ``Adaptive reduced-rank processing based on joint and iterative
  interpolation, decimation, and filtering,''
\newblock {\em IEEE Trans. Signal Process.}, vol. 57, no. 7, pp. 2503--2514,
  2009.

\bibitem{jiostap}
Rui Fa and R.C. de~Lamare,
\newblock ``Reduced-rank stap algorithms using joint iterative optimization of
  filters,''
\newblock {\em Aerospace and Electronic Systems, IEEE Transactions on}, vol.
  47, no. 3, pp. 1668--1684, July 2011.

\bibitem{ccmavf}
L.~Wang and R.C. de~Lamare,
\newblock ``Adaptive constrained constant modulus algorithm based on auxiliary
  vector filtering for beamforming,''
\newblock {\em Signal Processing, IEEE Transactions on}, vol. 58, no. 10, pp.
  5408--5413, Oct 2010.

\bibitem{Yao}
Yao Yu, A.P. Petropulu, and H.V. Poor,
\newblock ``Mimo radar using compressive sampling,''
\newblock {\em IEEE Journal of Selected Topics in Signal Processing}, vol. 4,
  no. 1, pp. 146--163, Feb 2010.

\bibitem{stapjidf}
Rui Fa, R.C. de~Lamare, and L.~Wang,
\newblock ``Reduced-rank stap schemes for airborne radar based on switched
  joint interpolation, decimation and filtering algorithm,''
\newblock {\em Signal Processing, IEEE Transactions on}, vol. 58, no. 8, pp.
  4182--4194, Aug 2010.

\bibitem{l1stap}
Zhaocheng Yang, R.C. de~Lamare, and Xiang Li,
\newblock ``L1-regularized stap algorithms with a generalized sidelobe canceler
  architecture for airborne radar,''
\newblock {\em Signal Processing, IEEE Transactions on}, vol. 60, no. 2, pp.
  674--686, Feb 2012.

\bibitem{vfap}
Jingjing Liu and R.C. de~Lamare,
\newblock ``Low-latency reweighted belief propagation decoding for ldpc
  codes,''
\newblock {\em Communications Letters, IEEE}, vol. 16, no. 10, pp. 1660--1663,
  October 2012.

\bibitem{l1cg}
Z.~Yang, R.C. de~Lamare, and X.~Li,
\newblock ``Sparsity-aware space-time adaptive processing algorithms with
  l1-norm regularisation for airborne radar,''
\newblock {\em Signal Processing, IET}, vol. 6, no. 5, pp. 413--423, July 2012.

\bibitem{alt}
R.C. de~Lamare and R.~Sampaio-Neto,
\newblock ``Sparsity-aware adaptive algorithms based on alternating
  optimization and shrinkage,''
\newblock {\em Signal Processing Letters, IEEE}, vol. 21, no. 2, pp. 225--229,
  Feb 2014.

\bibitem{Zheng}
Haipeng Zheng, S.R. Kulkarni, and H.V. Poor,
\newblock ``Attribute-distributed learning: Models, limits, and algorithms,''
\newblock {\em IEEE Trans. Signal Process.}, vol. 59, no. 1, pp. 386--398, Jan
  2011.

\bibitem{Bertrand}
A.~Bertrand and M.~Moonen,
\newblock ``Distributed adaptive node--specific signal estimation in fully
  connected sensor networks--part {II}: Simultaneous and asynchronous node
  updating,''
\newblock {\em IEEE Trans. Signal Process.}, vol. 58, no. 10, pp. 5292--5306,
  2010.

\bibitem{Rcdl3}
R.~C. de~Lamare and P.~S.~R. Diniz,
\newblock ``Blind adaptive interference suppression based on set-membership
  constrained constant-modulus algorithms with dynamic bounds,''
\newblock {\em IEEE Trans. Signal Proc.}, vol. 61, no. 5, pp. 1288--1301, March
  2013.

\bibitem{smce}
Tong Wang, R.C. de~Lamare, and P.D. Mitchell,
\newblock ``Low-complexity set-membership channel estimation for cooperative
  wireless sensor networks,''
\newblock {\em Vehicular Technology, IEEE Transactions on}, vol. 60, no. 6, pp.
  2594--2607, July 2011.

\bibitem{Cattivelli}
F.~Cattivelli, C.~G. Lopes, and A.~H. Sayed,
\newblock ``A diffusion {RLS} scheme for distributed estimation over adaptive
  networks,''
\newblock {\em Proc. IEEE International Workshop on Signal Processing Advances
  for Wireless Communications}, pp. 1--5, June 2007.

\end{thebibliography}

\end{document}